\documentclass[a4paper,12pt]{article}
\usepackage{amsmath,amssymb,amscd}
\usepackage{graphicx}
\usepackage{cite}

\usepackage{hyperref}
\providecommand{\url}[1]{#1}
\providecommand{\href}[2]{#2}

\newcommand{\figref}[1]{\ref{#1}}

\newcommand{\tref}[1]{(\ref{#1})}

\newcommand{\kred}{k_r}

\begin{document}

\begin{center}
   {\Large\textbf{What is the dimension of citation space?}}
 \\[0.5cm]
   {\large \href{http://www.jamesclough.net}{James~R.\ Clough}\footnote{james.clough09@ic.ac.uk}, \href{http://www.imperial.ac.uk/people/t.evans}{Tim S.\ Evans}\footnote{t.evans@ic.ac.uk}}
 \\[0.5cm]
    \href{http://complexity.org.uk/}{Centre for Complexity Science},
    Imperial College London,
 \\
    South Kensington campus, London, SW7 2AZ, U.K.
 \\ 30th July 2014   
%\tpreprint{ \\ (\texttt{\jobname.tex}  LaTeX-ed on \today ) }
 %\\  Key Words: Graph Theory, directed acylic graphs, citation networks
\end{center}

%\keywords{directed acyclic graph | causality | arXiv}
%\abbreviations{}

\begin{abstract}
Citation networks represent the flow of information between agents. They are constrained in time and so form directed acyclic graphs which have a causal structure. Here we provide novel quantitative methods to characterise that structure by adapting methods used in the causal set approach to quantum gravity by considering the networks to be embedded in a Minkowski spacetime and measuring its dimension using Myrheim-Meyer and Midpoint-scaling estimates.
We illustrate these methods on citation networks from the arXiv, supreme court judgements from the USA, and patents and find that otherwise similar citation networks have measurably different dimensions. We suggest that these differences can be interpreted in terms of the level of diversity or narrowness in citation behaviour.
\end{abstract}

% *****************************************************
\section*{Introduction}

Citation analysis has great potential to help researchers find useful academic papers \cite{H10b}, for inventors to find interesting patents \cite{AZK14}, or for judges to discover relevant past judgements \cite{M07}. It is not, however, enough to simply count citations, which can be made for a variety of reasons beyond an author genuinely finding a document useful \cite{B86,SR03,BD08}. To interpret the information encoded in a citation network we must also understand the structure of citation networks and the kinds of generative mechanisms which can create them.

These networks have a complex structure which is not easily described by any simple model and so are not easily characterised. The underlying processes which generate the network's structure are hard to observe directly but can be inferred from the network's structure itself by comparing networks to each other, or to models whose generating mechanisms we know. In order to compare two networks we need to be able to characterise their structure in ways relevant to the dynamics we are interested in. To begin to tackle this problem there has been much recent interest in trying to identify the statistical distribution of citation counts \cite{Price65, Price76, Redner1998, Tsallis2000, LLJ03, SR05a, RFC08, BD09c,SSA10, VG10, Eom2011}, and various other aspects of the topology of the citation network \cite{Newman2003a, Shanker2010, Bonato2014}.

When networks exist under some constraints, it is often possible to create new methods of characterising their structure which better take those constraints into account, as is well known for networks embedded in space \cite{Barthelemy2011,Expert2011,Daqing2011}. 
There is also interest in developing a geometric approach to studying network structure by finding a hidden space in which a network's nodes are embedded which then determines the network's structure\cite{Krioukov2010, Krioukov2012}. 

Citation networks are constrained in time, because authors can only cite something that has already been written\footnote{Occasionally this is not the case for real citation networks. 
For instance, two authors may share and cite each others work before either is published, leading to two papers which both cite each other, clearly forming a cycle. 
Such `acausal' edges are rare, making up less than 1\% of edges in all citation networks considered here, and so were removed from the network since many techniques used here assume that the network forms a DAG}.
This causal constraint prevents closed loops of directed edges in the graph, since all edges must point the same direction in time, and is the same constraint placed on causally connected events in physics.
They can therefore be represented as Directed Acyclic Graphs (DAG) where a directed edge goes from node A to node B represents document A having cited the document B. 

In this paper we approach these temporally constrained networks from a geometric point of view, except instead of an underlying manifold with a positive definite distance (such as usual Euclidean space of spatial networks, or a Hyperbolic space of \cite{Krioukov2010}) we consider a Lorentzian manifold; in particular, the simplest case which is Minkowski space, where points have a spatial position, but also a position in time. 
We will characterise the network's structure using tools from the causal set approach to quantum gravity which take a set of connected points and compare their causal structure to the causal structure of Minkowski space. 
In particular, we will use methods which estimate the dimension of a Minkowski space from its causal structure.

The rest of this paper is structured as follows. We will first introduce the causal set perspective of DAGs, seeing how they can be embedded in space and time, and the methods of estimating their dimension. In the second section we will adapt these methods for use on citation networks and test them on examples from academic papers, patents and court judgements. We will conclude by interpreting our results in terms of using dimension as a measure of citation diversity.

\section*{Dimension estimates for spacetime networks}

\begin{figure}[h!]
\centering
\includegraphics[width=0.25\textwidth]{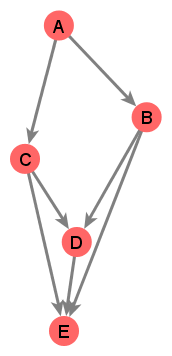}
\includegraphics[width=0.25\textwidth]{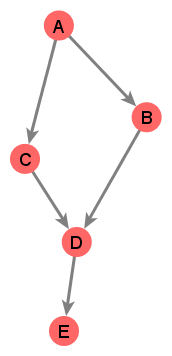}
\includegraphics[width=0.277\textwidth]{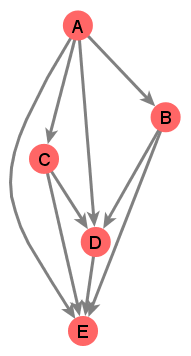}
\caption{\textbf{Left:} A DAG consisting of 5 points. A causal connection exists between any two nodes that can reach one another following edges along their direction. A and B, are causally connected by the edge (A, B); A and D are causally connected by the path \{A, B, D\}; but B and C are not causally connected. The interval [A, E] contains all of the nodes in this graph since B, C and D are all causally connected to A in one direction, and E in the other direction. Like most citation networks, this graph is neither transitively complete, or transitively reduced.
\newline
\textbf{Centre:} The transitive reduction of this graph. All edges except those required to keep all of the causal relations of the DAG have been removed. For example, the edge (B, E) is not required since B and E are already causally related by the path \{B, D, E\}. No causal connections have been created or destroyed. The edges that remain after TR are called `links' in the causal set literature, but elsewhere can be called `covering relations' or `nearest neighbour edges'. In the rest of this paper we will draw networks after TR as it makes the structure of the network easier to see, but does not break any causal connections.
\newline
\textbf{Right:} The transitive completion of this graph. All pairs of causally related nodes now have an edge drawn between them, or alternatively, all edges implied by transitivity are added. Spacetime networks are, by construction, always transitively complete.}
\label{fig:example}
\end{figure}

In the causal set approach to quantum gravity, spacetime is seen as a set of discrete points with causal relation, called a causal set whose structure approximates the continuous space we perceive. We direct the reader to \cite{Bombelli1987,Brightwell1991,Dowker2006} for more details. 

We will consider only the simplest spacetimes, $D$ dimensional Minkowski spacetimes of one time dimension and $(D-1)$ spatial dimensions.\footnote{It is possible to define other similar networks, such as a cube-space\cite{Bollobas1988}, or a spacetime network using a more complicated spacetime\cite{Krioukov2012}. Minkowski spacetime is the simplest, being defined by just one parameter $D$, the measurement of which we will use to characterise the network's structure. Furthermore, all other Lorentzian manifolds can be approximated, locally, as Minkowski space.}
To create a causal set which approximates the structure of Minkowski space, we begin by randomly and uniformly scattering points in Minkowski space by randomly assigning each point an associated time $t$ and spatial co-ordinates $x_i$. 
Two points are causally connected if and only if the differences in their co-ordinates satisfy:
\begin{equation}
(\Delta t)^2 > \sum_i (\Delta x_i)^2
\label{eq:minkowski}
\end{equation}
which is to say their separation in time is larger than their separation in space, using the speed of light to convert between the units of space and time (equivalently we choose units where the speed of light is equal to $1$). If this relationship is satisfied we then say that the point with the larger/smaller $t$ coordinate is in the future/past lightcone of the other. 
In special relativity it is this relationship that defines whether two events in spacetime can causally affect one another. The direction of the edges is determined by the causal/temporal ordering as given by the ordering of the time coordinates, and provides a uniquely defined causal relationship.
To translate this structure into the language of networks, we say each point is a node, and we add edges between nodes which are causally connected, i.e.\ satisfying \tref{eq:minkowski}, with a direction reflecting the flow of time.  We will use the convention that all edges point backwards in time.  Such a network is necessarily a DAG because all of the edges point the same directed in time.

An \textbf{interval} [A, B] in a DAG is the set of nodes which can be reached from A (are in its causal past) in one direction, and from B in the other direction (in its causal future)\cite{Henson2010} as in figure~\ref{fig:example}. The dimension estimates used here are defined on an interval in a DAG.

To illustrate these estimates we will use a simple network model involving randomly scattering points in an interval in Minkowski space. We first create two extremal points with time co-ordinates of $0$, and $1$ respectively, and all spatial co-ordinates of $0$. 
We then add more nodes to the network by assigning a random time co-ordinate between $0$ and $1$, and random spatial co-ordinates between $-0.5$ and $0.5$ and allowing this node to be in the network if it has edges to the two extremal nodes and so lie within the interval such that $G_D (N)$ is a network created by this process with $N$ nodes, which have are described by $D$ coordinates.
We will refer to these networks as \textbf{spacetime networks} though they are also known as cone spaces in the mathematics literature \cite{Bollobas1991}.

The number of spatial dimensions will determine the structure of the graph this process creates. Extra spatial dimensions add further terms to the summation on the right hand side of \tref{eq:minkowski} and make it less likely that two points are connected.
So if we were to forget about the space and time coordinates of each point, it would be possible to estimate the number of spatial dimensions by looking only at the network's structure.
We will use two such methods: the Midpoint-Scaling dimension estimate, and the Myrheim-Meyer dimension estimate.

\subsection*{Midpoint-Scaling Dimension}

When nodes are uniformly and randomly scattered in a space, the number of points in a region is proportional to the volume, $V$, of that region\cite{Meyer1988}. 
In a Minkowski space, the longest path through an interval corresponds to the geodesic through the continuous spacetime limit (the Myrheim length conjecture) \cite{Thompson2003, Myrheim1978}. 
This means that in an interval the length of the longest path, $L$ is proportional to the time difference between the starting and ending nodes.\footnote{The causal structure of Minkowski space and therefore the structure of the spacetime network is invariant under Lorentz boosts, so any interval can be transformed such that the starting and ending nodes have the same spatial coordinates, so the length of the geodesic in the continuum limit is simply the difference in the node's time coordinates.}
We then expect, in a D-dimensional Minkowski space that $V(L) \thicksim L^D$. Knowing how the size of an interval scales with its height allows the dimension to be inferred.

The Midpoint-Scaling dimension \cite{Brightwell1991} measures how the size of two subintervals scale with the size of a larger interval between two nodes. The two subintervals of interval [A, B] are [A, C] and [C, B], which have populations $N_1$ and $N_2$. The midpoint, C, is the node on the longest path such that smaller population of $N_1$ and $N_2$ is maximised.

\begin{figure}[h!]
\centering
\includegraphics[width=0.8\textwidth]{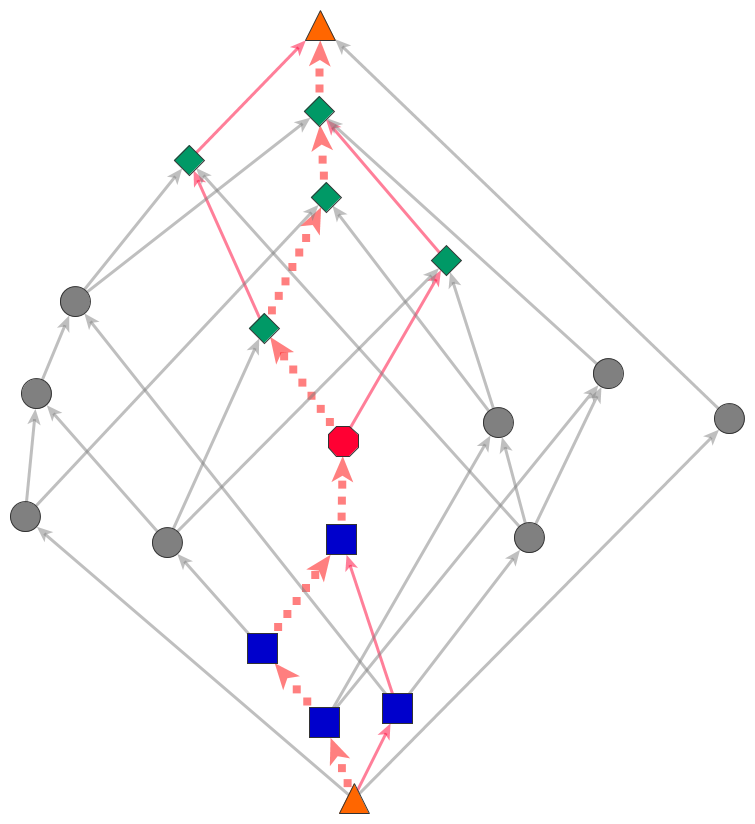}
\caption{The Midpoint-Scaling dimension in a 2-D spacetime network. The longest path through the interval (defined by the triangular nodes) is shown, and its midpoint is the octagonal node. The diamond nodes are those that lie within a subinterval, from the midpoint to the upper extremal point of the network, and the square nodes lie in the lower subinterval. In a 2D network we expect the number of nodes lying in these subintervals (the diamonds and squares) to be approximately half of the total population of the whole network. 
For simplicity, only the the essential links, those remaining after TR are drawn here. It is only the essential links that matter for these dimension estimates, since it is only the causal structure that determines them.}
\label{fig:MPSD}
\end{figure}

Since [A, C] and [C, B] each have around half the height of [A, B] we can estimate the manifold dimension of this interval using $N_1 \simeq N_2 \simeq \frac{N}{2^D}$. This is illustrated in figure~\ref{fig:MPSD}.

\subsection*{Myrheim-Meyer Dimension}
An \textbf{n-chain} in a DAG is a sequence of $n$ nodes which are all causally connected to each other. 
When points are placed at random with uniform probability density in spacetime in an interval the expected number of $n$-chains $S_n$, is known to be \cite{Myrheim1978, Meyer1988, Reid2003}
\begin{equation}
 \langle S_{n} \rangle = \frac{N^{n} \Gamma (D/2) \Gamma (D) \Gamma(D + 1)^{n-1}} { 2^{n-1}n\Gamma(nD/2)\Gamma((n+1)D/2) }
 \label{eq:MMD}
\end{equation}
where $D$ is the dimension of the Minkowski spacetime and $\Gamma(z)$ is the standard Gamma function.

Given a DAG we can simply count the number of chains and numerically find an estimate for the dimension using this formula. Estimates for dimension can be made using chains of any length $n$, but there are significantly larger number of 2-chains (just a pair of causally connected points) and so these produce the most accurate estimation of dimension. The expected number of 2-chains is simply
\begin{equation}
 \frac{\langle S_{2} \rangle}{N^2} \equiv f(D) =  \frac{\Gamma(D+1)\Gamma{(D/2)}} { 4\Gamma(\frac{3}{2}D) }
\end{equation}
For a given interval, the left hand side of this equation can be measured, and the right hand side, $f(D)$ is a monotonically decreasing function, so we can estimate $D$ by inverting it numerically.

\begin{figure}
\centering

\begin{tabular}{ c c  }
 \includegraphics[width=0.4\textwidth]{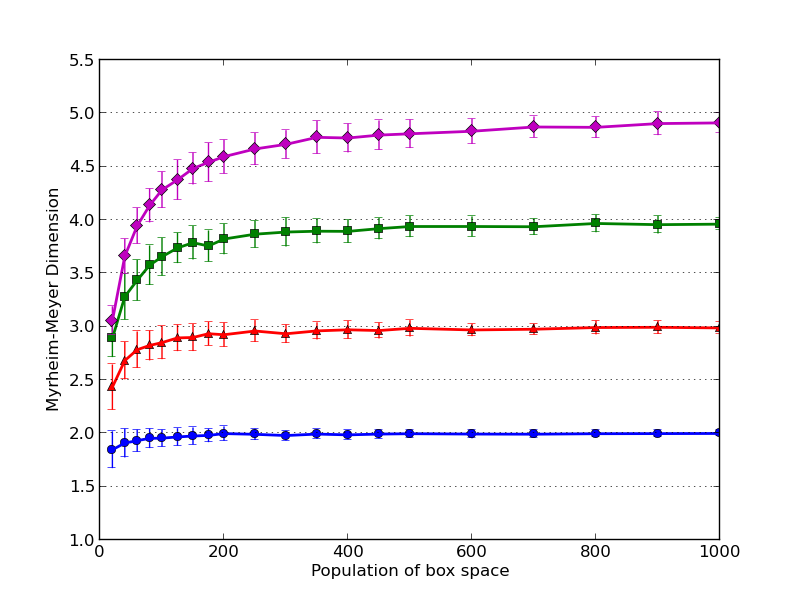} &
 \includegraphics[width=0.4\textwidth]{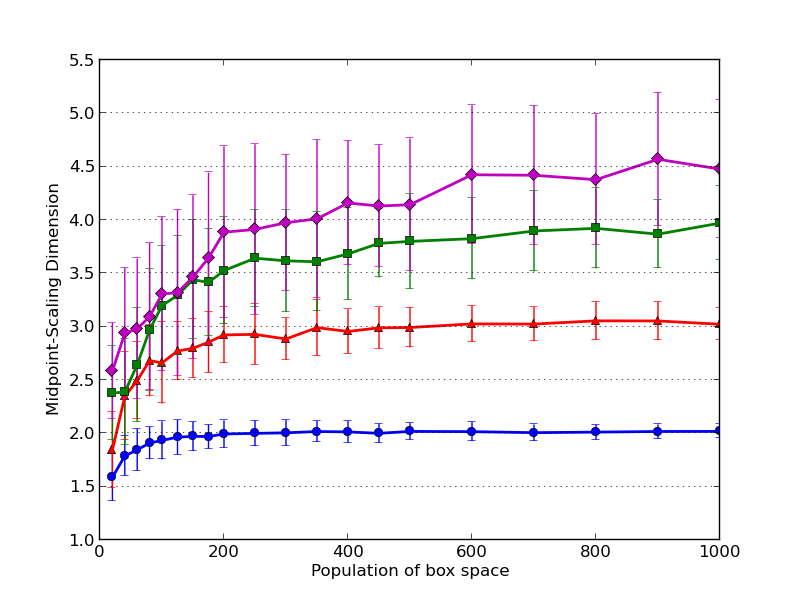} \\
\end{tabular}

\caption{The Myrheim-Meyer dimension (left column) and Midpoint-scaling dimension (right column), against the number of points in the graph, averaged over 100 random spacetime networks $G_D(N)$ for $D$ equal to 2, 3, 4 and 5.
The estimates' convergence from below to the correct dimension is also seen in \cite{Reid2003}. Error bars show the standard deviation of the estimated dimension. Errors are larger for higher dimension, but smaller as the size of the space grows.}
\label{fig:box}
\end{figure}

\subsection*{Reduced degree}
It is also possible to estimate the dimension of a spacetime network by comparing the average in/out degree of a node before and after Transitive Reduction (TR).
TR is an operation on directed graphs which removes all edges implied by transitivity, the result of which is uniquely defined if the graph is acyclic \cite{Aho1972} (see figure~\ref{fig:example}). We will call the degree after TR the \textbf{reduced degree}, $\kred$.
Taking figure~\ref{fig:MPSD} as an example, consider the triangular node at the bottom of the diagram. It's degree before TR is $N=20$ and its degree after is $\kred = 4$ (as shown by the 4 remaining links).

We show in appendix A that for a two-dimensional spacetime network the distribution of reduced degrees $\kred$ is proportional to the unsigned Stirling numbers of the first kind. For large $N$ the degree distribution is roughly Poissonian with a mean of $\ln(N)$. For other dimensions the expected reduced degree is roughly $\kred^{\frac{D}{D-2}}$ \cite{Bombelli1987}\footnote{See appendix A for derivation}.
However, we found that the dimension estimate given by this method does not display the consistency of the other two methods described here when used on citation networks. This is primarily because in a given citation network nodes which have the same degree can have reduced degrees which differ by more than an order of magnitude, as shown in appendix A.
In \cite{Clough2013a} we suggest that the reduced degree of a node reveals particular properties of the paper it represents, and given such variation this method is too noisy to use as a way of characterising the network as a whole.

\section*{Estimating the Manifold Dimension of citation networks}
\subsection*{Adapting the methods to citation networks}

The methods described above are designed to estimate the dimension of the spacetime in which the nodes of a DAG are randomly scattered. DAGs which represent citation networks do not originate from points scattered in a Minkowski space, and so there is no original `dimension' for us to estimate.

Despite this, these algorithms to estimate dimension can be applied to any DAG, and a result can be obtained. However, our interpretation of this result does have to change. 
We are now no longer investigating the properties of a space in which the nodes are embedded, but instead just characterising the DAG's structure in a way that is \emph{analogous} to embedding it in some Minkowski spacetime. 
We do not claim that citation networks actually have the same structures as the spacetime networks described above, only that these tools are useful characterisations of different citation networks.

Some work is needed to adapt these dimension estimators to use on citation networks  because citation networks do not necessarily share some of the particular properties of the spacetime networks used in these estimators.

Firstly, the spacetime networks are constructed to be an interval, that is there is only one `start', or `source' node (with zero in-degree) and one `end' or `sink' node (with zero out-degree), both of which are reachable from any node in the network. This is almost always not the case for citation networks.  So instead of estimating the dimension of the whole citation network, we look at many small intervals within the citation network and apply the estimators to these intervals. To find intervals we choose two nodes uniformly at random from the network, and if an interval exists between them we estimate its dimension, otherwise we ignore this pair of nodes.  We then plot the population of the interval against its estimated dimension.

Secondly, the spacetime networks are always \textbf{transitively complete}. That is if node A is in the future lightcone of B, and B is in the future lightcone of C then A is necessarily in the future lightcone of C.  In the network the edges (A, B) and (B, C) imply (A, C). In citation networks this constraint is not present since if an author cites a paper, they do not also have to cite its entire bibliography.
A consequence of this is that there is no distinction in spacetime networks between edges and causal connections, but in citation networks they are different. So in our implementation of the Myrheim-Meyer dimension estimator we seek to count chains of causally connected nodes and not just edges. To do this we first transitively complete the network \cite{Aho1972} (adding edges between any two nodes if there is a path between them) before counting the 2-chains which are now just the edges.

\subsection*{Data}
To test these dimension estimates we used citation networks from academic papers, patents and court judgements.
The academic citation networks are from subsections of the arXiv online research paper repository, from the citation network visualiser  paperscape \cite{paperscape}.
The citation network is separated out into the subsections of the arXiv, and each consists of the citations from one paper in that subsection to another also in that subsection. Here we will look at the `high energy theory', `high energy phenomenology', `astrophysics', and `quantum physics' sections, labelled by their tags on the arXiv, \texttt{hep-th}, \texttt{hep-ph}, \texttt{astro-ph}, \texttt{quant-ph} respectively. Their sizes range from around 20,000 to around 120,000 nodes and stretch in time from 1991 to 2013.

Since patents must cite other patents that contain `prior art' they also form a citation network. We use data derived from patents registered in the USA between 1975 and 1999 \cite{Hall2001} and in total there are around 4,000,000 patents.

Court decisions also cite previous decisions as precedent so form a citation network. We will analyse the network formed by all decisions and citations made by the US Supreme Court from its inception in 1754 to 2002 \cite{Fowler2008}, in total around 25,000 nodes.
Further discussion of these particular datasets is available in our previous paper \cite{Clough2013a} and our datasets will be made available on figshare \cite{figshare_dimension}.

%*****************************************************************************
\section*{Discussion and Interpretation}

\begin{figure}[!h]
\centering
\includegraphics[width=0.49\textwidth]{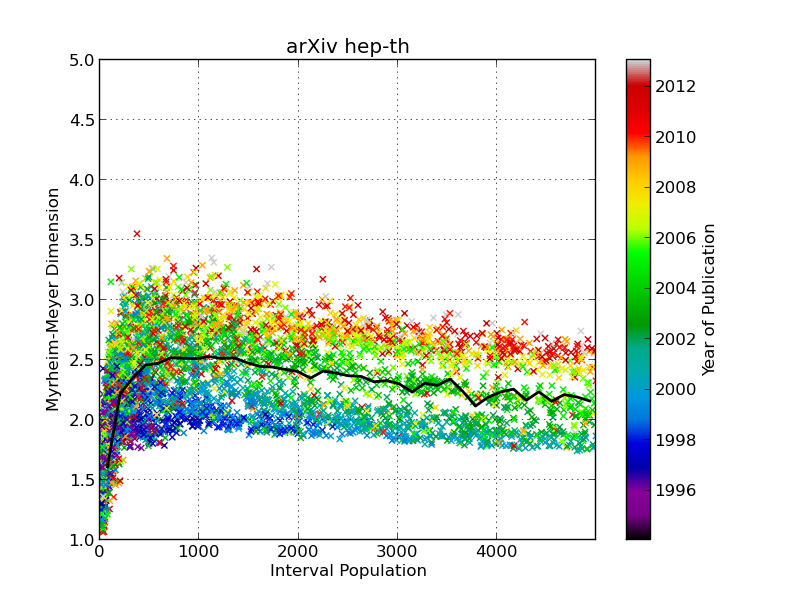}%}
\includegraphics[width=0.49\textwidth]{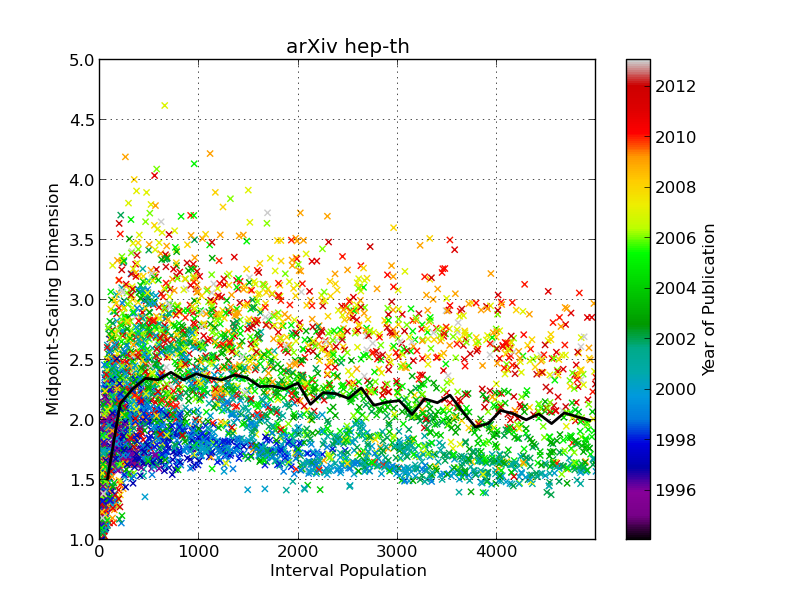}%}
\caption{The Myrheim-Meyer (left) and Midpoint-scaling (right) dimensions for the \texttt{hep-th} citation network appears to settle at a value around 2 for large intervals}
\label{fig:hep-th}
\end{figure}

\begin{figure}[!h]
\centering
\includegraphics[width=0.49\textwidth]{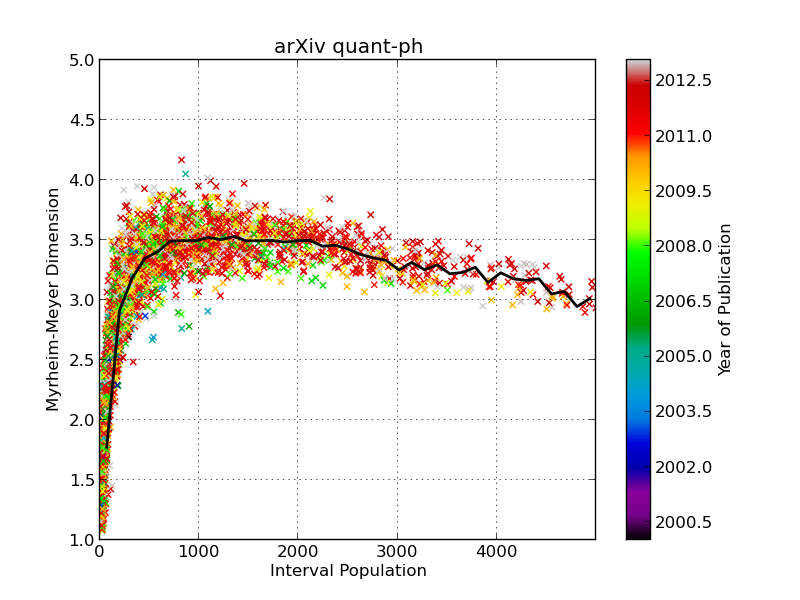}%}
\includegraphics[width=0.49\textwidth]{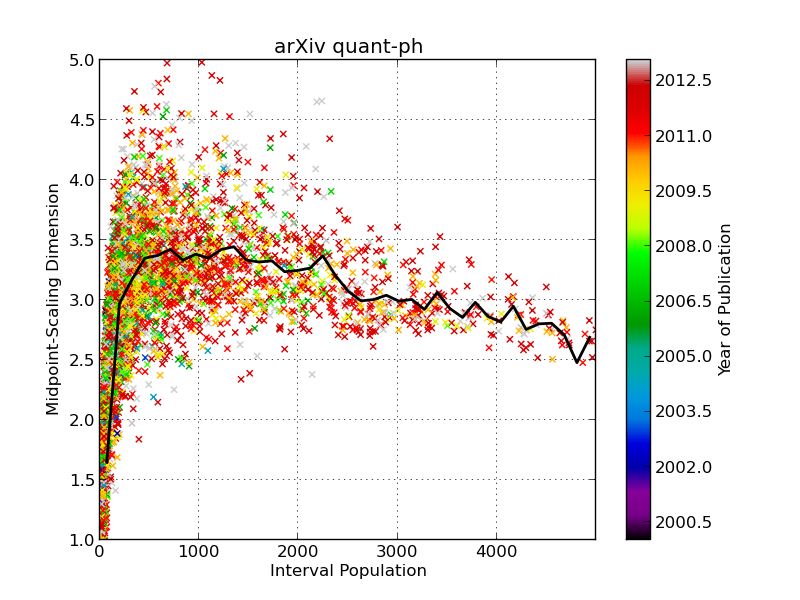}%}
\caption{The Myrheim-Meyer (left) and Midpoint-scaling (right) dimensions for the \texttt{quant-ph} citation network appears to settle at a value around 3 for large intervals}
\label{fig:quant-ph}
\end{figure}

\begin{figure}[!h]
\centering
\includegraphics[width=0.49\textwidth]{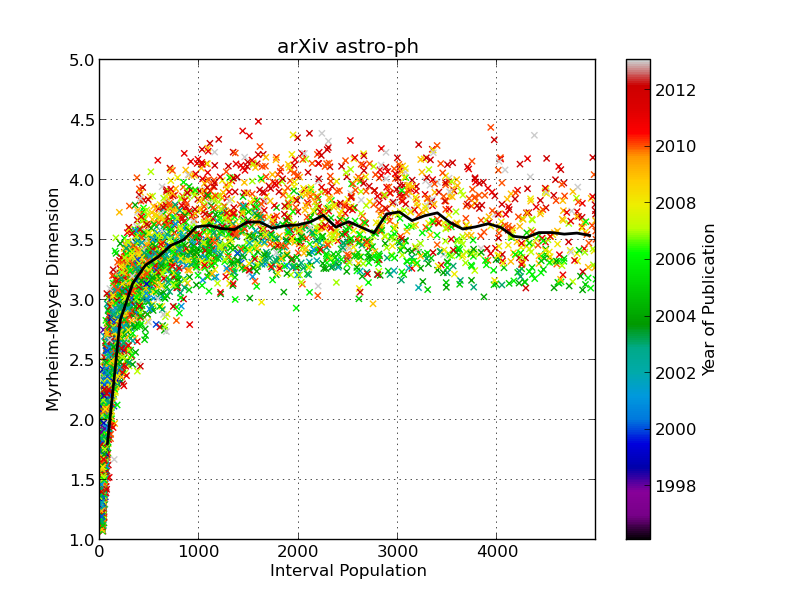}%}
\includegraphics[width=0.49\textwidth]{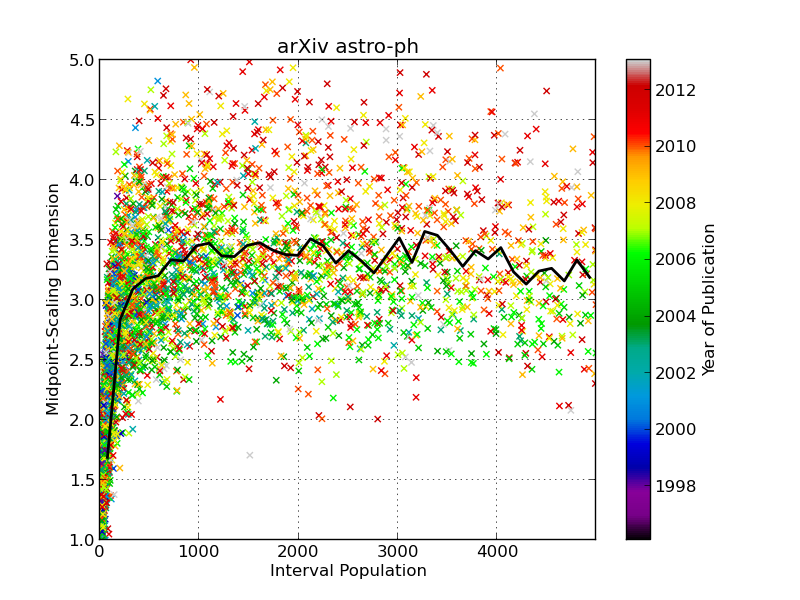}%}

\caption{The Myrheim-Meyer (left) and Midpoint-scaling (right) dimensions for the \texttt{astro-ph} citation network appears to settle at a value around 3.5 for large intervals}
\label{fig:astro-ph}
\end{figure}

\begin{figure}[!h]
\centering
\includegraphics[width=0.49\textwidth]{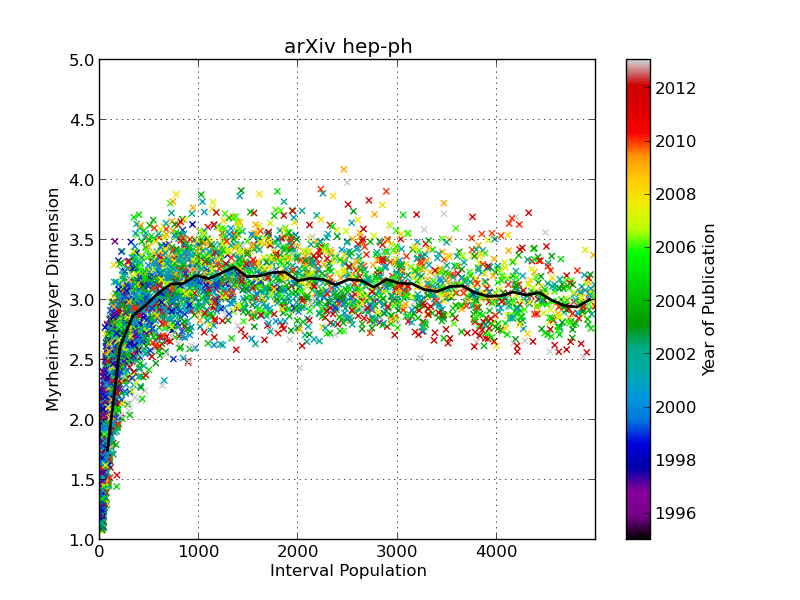}%}
\includegraphics[width=0.49\textwidth]{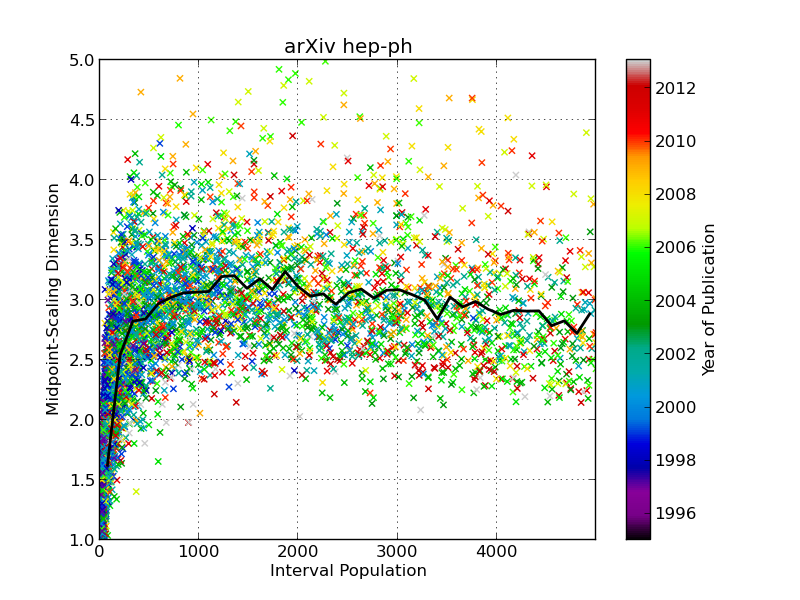}%}

\caption{The Myrheim-Meyer (left) and Midpoint-scaling (right) dimensions for the \texttt{hep-ph} citation network appears to settle at a value around 3 for large intervals}
\label{fig:hep-ph}
\end{figure}

Figures \figref{fig:hep-th}, \figref{fig:quant-ph}, \figref{fig:astro-ph} and \figref{fig:hep-ph} scatter plots for each arXiv section, plotting the population of an interval against its estimated dimension. Each point in coloured by the publication date of the last node in the interval allowing us to see how the estimated dimension changes as more papers are added to the citation network.
\newpage

\clearpage

\clearpage

\begin{figure}
\centering
\includegraphics[width=0.49\textwidth]{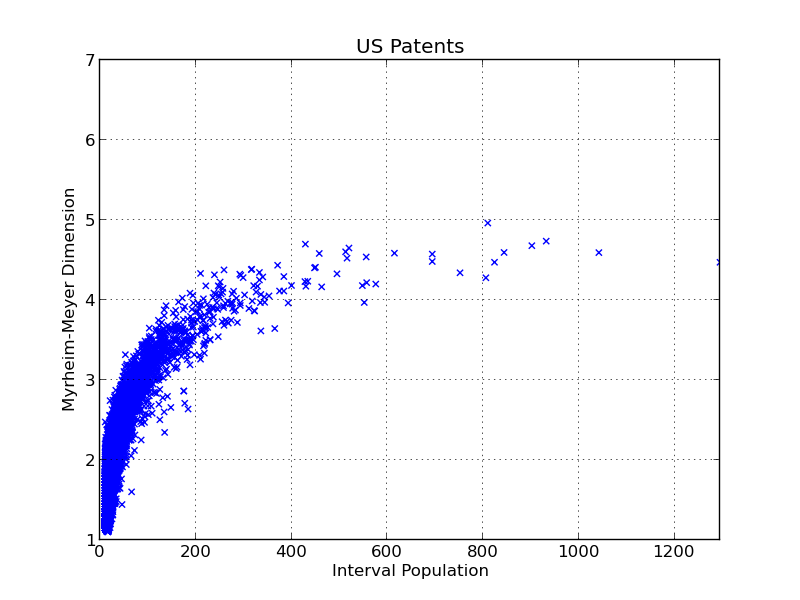}%}
\includegraphics[width=0.49\textwidth]{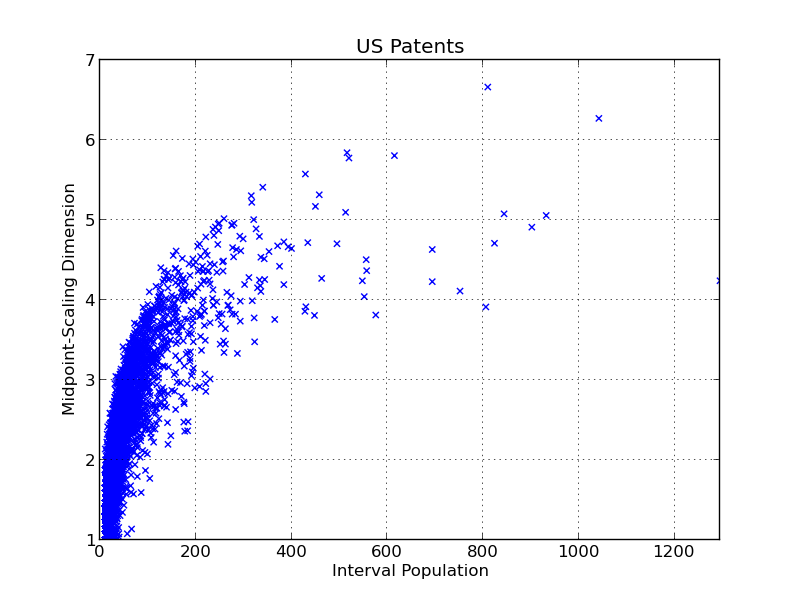}%}

\caption{The Myheim-Meyer (left) and Midpoint-scaling (right) dimensions for the patent citation network. In this citation network larger intervals are much rarer than in the others, as a large interval usually contains many different paths from that starting node to the ending node, which is rare in patent citation networks.}
\label{fig:patent}
\end{figure}

%heterogeneous
It is immediately clear from the differing shapes of the histograms in figures~\figref{fig:hep-th} to~\figref{fig:USSC} that there are structural differences in the citation networks analysed here which have been revealed by these dimension estimates. 
In general there is a large spread of measured dimensions suggesting structural heterogeneity unlike the homogeneous Minkowski space. This is  unsurprising given that citation networks show high levels of clustering  and usually contain many different communities with strong intra-community links but weak inter-community links \cite{Newman2003a}.

In all four arXiv citation networks the two plots, for Myrheim-Meyer dimension and Midpoint-Scaling dimension, show similar shapes, and converge on a consistent dimension value for large interval sizes. 
For networks generated from scattering points in a Minkowski space there is an underlying dimension being estimated and so it is reasonable to expect independent methods to agree. 
This is not obviously the case in other networks so it is encouraging to see consistency between the two methods in real social networks.

Crudely, the `dimension' of the \texttt{hep-th} network appears to be around 2, and the \texttt{hep-ph} network around 3, \texttt{astro-ph} around 3.5, and \texttt{quant-ph} also around 3.
We note that each of the individual arXiv citation networks, containing only intra-section links are themselves sub-networks of the larger arXiv citation network. They have significantly different estimated dimensions strongly suggesting that the arXiv citation network is structurally heterogeneous, with its different communities having measurably different citation behaviours. We suggest that these estimates could provide a novel method of measuring these differences in other large, heterogeneous citation networks, or DAGs representing other systems, and can identify relevant subgroups even without externally applying labelling of the nodes.

\subsection*{Similar causal constraints give similar structure}
\begin{figure}[!ht]
\centering
\includegraphics[width=0.49\textwidth]{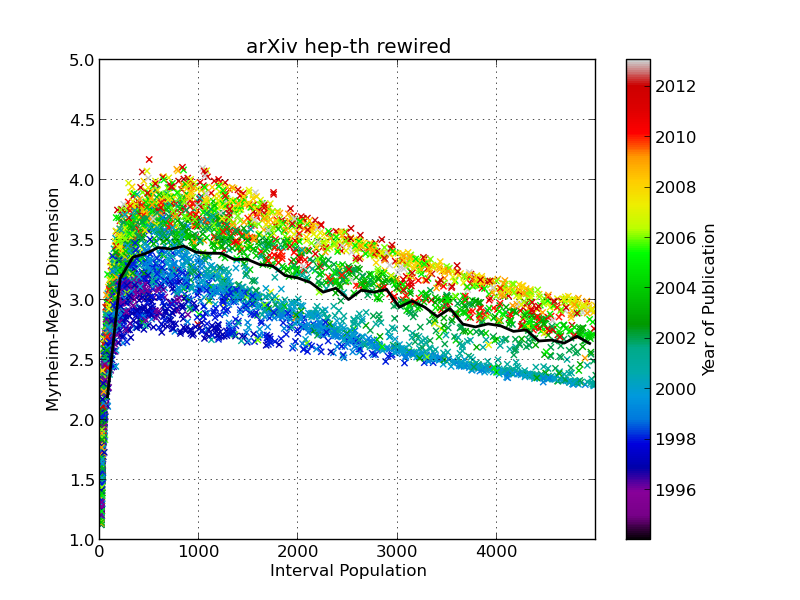}%}
\includegraphics[width=0.49\textwidth]{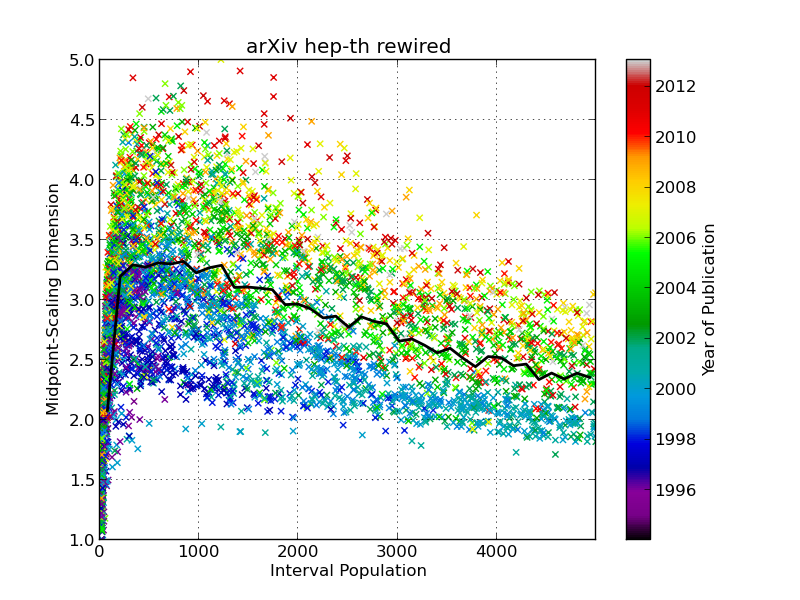}%}
\caption{The Myrheim-Meyer (left) and Midpoint-scaling (right) dimensions for the rewired \texttt{hep-th} citation network. The $~10^5$ edges have been rewired randomly $10^7$ times, so all structure other than the degree distribution and causality constraints has been removed.}
\label{fig:hep-th_rewired}
\end{figure}

Citation networks are under causal constraints which impose some structure. 
We can see the effect of this structure by rewiring the edges of the network but maintaining the causal constraints. 
This is done by taking two edges, [A, B] and [C, D] and rewiring them to [A, D] and [C, B] if both of the new edges respect causality, thereby retaining the original in, and out degrees of each node and ensuring the network remains a DAG. 
Figure \figref{fig:hep-th_rewired} shows that after all structure other than the causal constraints and in and out degree of each node in a network is removed, the dimension estimate plots give a significantly different result to the original \texttt{hep-th} network suggesting some other structure is involved in determining the estimated dimension and it is not an expected feature of a random network under the same constraints.

%%%

\subsection*{Null models}

\begin{figure}[!ht]
\centering
\includegraphics[width=0.49\textwidth]{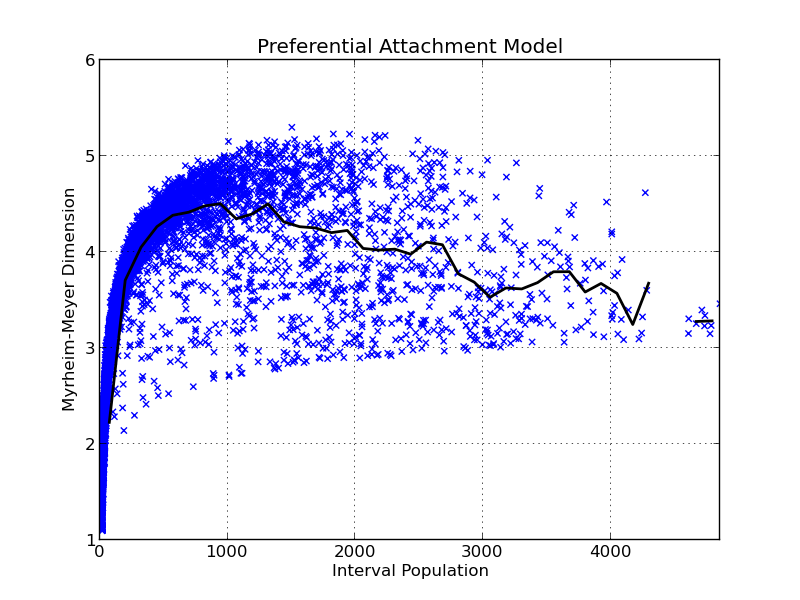}%}
\includegraphics[width=0.49\textwidth]{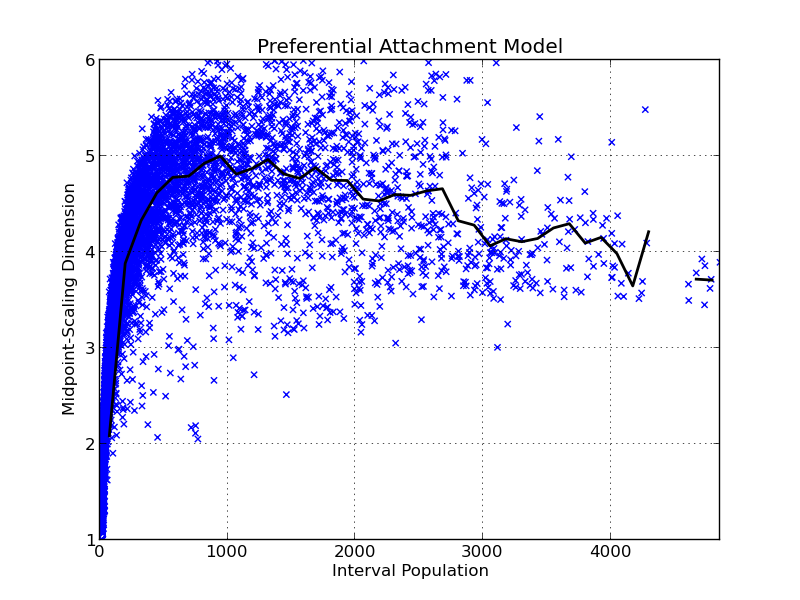}%}
\caption{The Myheim-Meyer (left) and Midpoint-scaling (right) dimensions for the Price model network, with the same size and average degree as the \texttt{quant-ph} citation network. The spread of values is much larger than in the real citation network and the dimension estimate is higher, illustrating how these dimension estimates can show differences in structure.}
\label{fig:ba}
\end{figure}

Here we further investigate the extent to which causal constraints and degree distribution determine estimated dimension with a simple null model.

We generate a network with the same degree distribution as the \texttt{quant-ph} network using the simple cumulative attachment model for citations\footnote{We begin with a small number of nodes connected in a line. We add nodes one by one, and when a node is added it attaches $\langle k_{in} \rangle$ edges to existing nodes, where $\langle k_{in} \rangle$ is the mean in degree in the network whose degree distribution we are replicating. With probability $p$, edges attach preferentially, that is, proportionally to nodes according to their current in-degree, and with probability $1-p$ they attach randomly. By manually tuning $p$, we can create a network with a very similar degree distribution to a real citation network. In this instance, $p=0.6$.} due to Price \cite{Price76}
%\cite{Barabasi1999a}
and measure its dimension.

Figure~\figref{fig:ba} shows that we can easily create a scale-free network with the same degree distribution as a citation network does not look the same according to these dimension estimates.

\subsection*{Differing citation behaviour}

\begin{figure}
\centering
\includegraphics[width=0.49\textwidth]{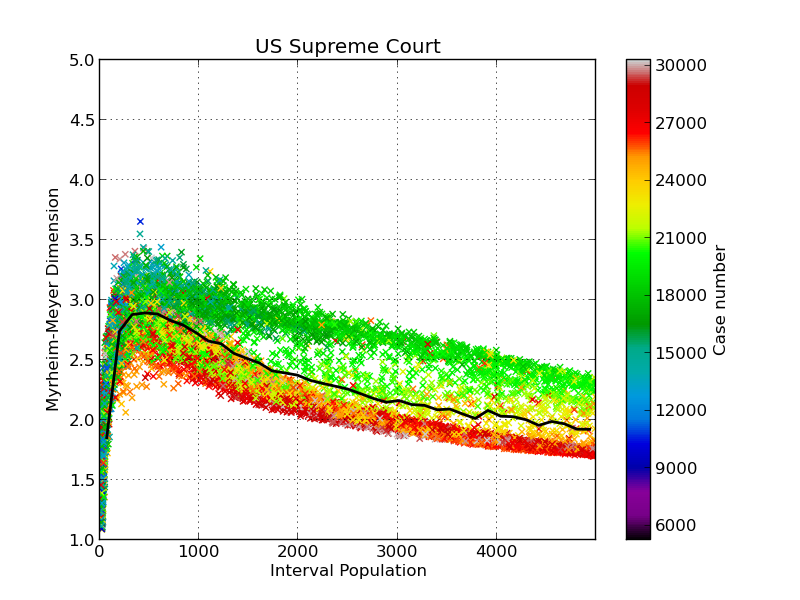}%}
\includegraphics[width=0.49\textwidth]{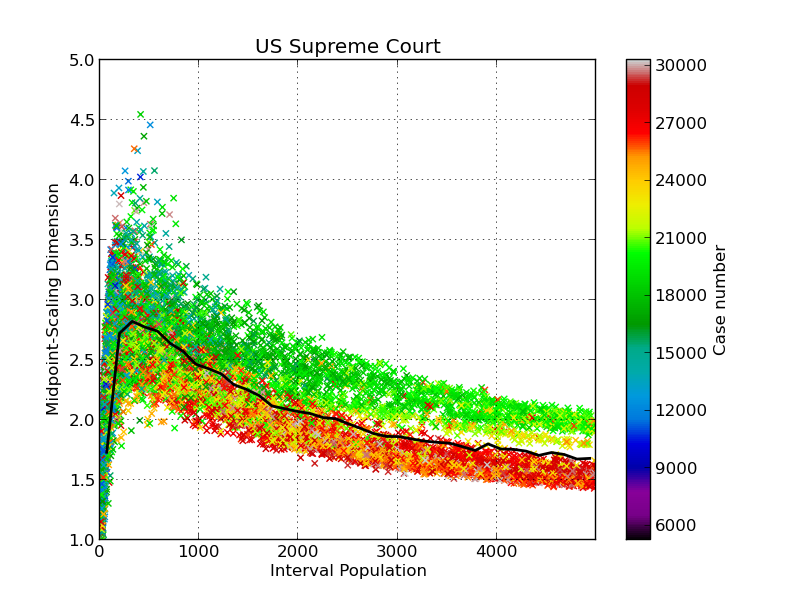}%}
\caption{The Myrheim-Meyer (left) and Midpoint-scaling (right) dimensions for the US Supreme Court citation network.}
\label{fig:USSC}
\end{figure}

Citation networks from outside academia illustrate different behaviour. The US patent network's plot is much sparser for larger intervals, and almost all measured intervals were very small. 
%This network, as discussed in \cite{Clough2013a}, has much lower clustering than the others discussed here and fewer of its edges are implied by transitivity. #
For large intervals the measured dimension is around 5, much larger than the arXiv citation networks.

% dependence on interval size - increasing
The plot for the US Supreme Court citation network has a different shape to all the others we investigated. In the arXiv and patent citation networks we see a slow growth in estimated dimension as interval size increases. 
% dependence on interval size - decreasing
The US Supreme Court network, figure~\figref{fig:USSC} seems to show the opposite effect. Small intervals have a higher dimension estimate, and dimension falls as interval size increases. Our suggestion is that this effect is caused by this network stretching over an unusually long time period (it covers all judgements made in the Supreme Court since 1754). In the same way that a large, thin plane appears three-dimensional on length scales much smaller than the plane's thickness, but two dimensional on length scales much larger than this thickness, this network may also appears to have a different estimated dimension on different time scales. 
We also note that newer cases (those with a larger case number on the colour bar) have a lower measured dimension than older cases which is again the opposite trend to that in the arXiv citation networks.
This temporal heterogeneity is another aspect of citation network structure which can be revealed by measuring dimension.

\subsection*{Measuring Citation Diversity}
To find an interpretation for these results, we can look at simple examples of low, and high dimensional networks. A network with dimension 1 (no spatial dimensions, just a time dimension) would be a single line, where nodes form an unbroken chain of edges, each linking the previous node. 

As the number of spatial dimensions grow it is more likely that two nodes do not have a causal relationship, and so do not link to each other\footnote{In the language of causal sets, the \emph{ordering fraction} decreases.}. 
Furthermore in any spatio-temporal model of citation networks the number of spatial dimensions corresponds to the number of coordinates required to parametrise a paper.
It is appealing then to interpret a small dimensional citation network as a more narrow field, where most of the papers are causally related to most others, and any paper can be described by a small number of paramters, corresponding to a small number of different areas of study.

\begin{figure}[h!]
\centering
\includegraphics[width=0.1\textwidth]{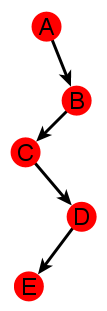}
\includegraphics[width=0.25\textwidth]{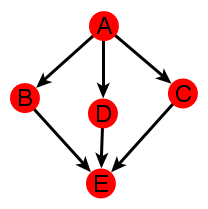}
\includegraphics[width=0.25\textwidth]{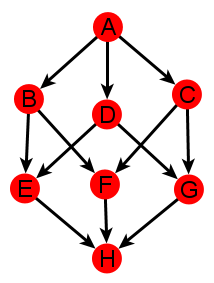}
\caption{From left to right, simple DAGs embeddable in 1, 2 and 3 dimensional Minkowski spacetimes.}
\label{fig:dimension_examples}
\end{figure}

A large dimension would then correspond to a more diverse field, where many independent authors can cite the same paper without citing each other and each paper requires a large number of parameters to be described.
Our high-dimensional networks, such as the astrophysics section of the arXiv can then be interpreted as being more diverse in terms of citation behaviour than high energy physics section, and patents more diverse than physics papers or court judgements.

\section*{Discussion}
% general comments about effectiveness of measures
In this paper we have illustrated the effectiveness of manifold dimension estimates as novel ways of characterising networks which form DAGs, and  in particular networks constrained by causality. We have shown that in citation networks, the Midpoint-Scaling dimension and Myrheim-Meyer dimension estimators show strong agreement and highlight important differences in the causal structure.

For instance the two particle physics sections of arXiv, \texttt{hep-th} and \texttt{hep-ph}, are similar in many ways but clearly differ in the dimension measures which quantify how `broad' or `narrow' the citation behaviour of authors in these fields is.
Furthermore, these methods can be used as a way of differentiating citation networks. Given two intervals, one from the \texttt{hep-th} network, and the other from \texttt{hep-ph} we can estimate their dimensions using these methods and we could deduce which section of the arXiv those papers came from, without knowing their authors or content, using only the topology of the citation network.

The message here is that citation networks from different areas of study have measurably different citation behaviours, and that this is potentially useful information we can extract which is potentially of interest to any scientists who want to improve their use of bibliometric measures as an aid to research.

% other ways of talking about dimension
\subsection*{Why use Minkowski space?}
The dimension of a general network is a concept which has been considered before, with differing meanings; for example the number of parameters in a model \cite{Freeman1983}, the time a random walker takes to return to its starting position \cite{Bilke2001}, fractal box counting dimension in Euclidean space \cite{Song2007}, or the number spatial (but not temporal) dimensions measured using influence regions \cite{Bonato2014}.

However causal constraints are a key feature of citation networks and DAGs and it is essential that such a constraint is taken into account when analysing these networks \cite{Clough2013a}. This is why it was important to develop methods which include time as dimension which is not the same as the spatial dimensions. The dimension measures we use explicitly involve time and take the the causal constraints of citation networks into account, recognising that in-edges (being cited) and out-edges (citing someone else) are fundamentally different things. While in a purely spatial embedding of a network, one might reasonably suggest that the probability of an edge existing between two nodes will always increase if the distance between them decreases, this is not necessarily the case in time. The information in one paper takes time to reach other researchers, and this delay is larger if researchers work in separate field. 

In the arXiv citation network, we found that the time difference of a citation between two papers in the same arXiv subsection was 1.6 years, and for papers in different subsections 2.1 years, suggesting that citations to papers which are less similar (or further away in some abstract spatial representation) are more likely to span larger time intervals, reflecting the finite speed of information propagation.

The primary reason for our choice of Minkowski spacetime for these dimension estimates over some other spacetime is that it is one of the simplest choices to make, defined by a single parameter $D$. The scope of this work was to characterise different network structures by measuring spacetime dimension and not to find an actual embedding of the network into that spacetime by assigning coordinates to each point. 

This would be much more difficult to do for spacetimes which are described by more than just one parameter (eg. for curved spacetimes). For studies of networks embedded in other types of space, for causal sets we refer the reader to \cite{Dowker2006,Reid2003,Bolognesi2014} and for growing network models in hyperbolic spaces to \cite{Krioukov2012, Ferretti2014}. 

There are suggestions in the literature, that it may be possible to find better correspondences between particular networks and spaces. The de Sitter spacetime studied in \cite{Krioukov2012} can give the same fat-tailed citation distribution as the simple Price model \cite{Price65} but the development of other network measures is needed before we can truly say that there is a correspondence between citation networks and these kinds of spaces. Indeed, using other causally aware measures on citation data reveals important new features in real citation network data \cite{Clough2013a,Goldberg2014} which are not present in generic preferential attachment models. 

One particular feature noted in \cite{Clough2013a} is that only small number of papers published shortly before the referencing paper are needed to define the causal structure of a real citation network.  These are the essential links, i.e. the citations left after transitive reduction.

We note that other dimension estimators exist for networks embedded in Minkowski spaces. We have tried using the reduced degree, but as explained in the appendix it is not a feasible method.
A recently published method not implemented here, but potentially appropriate for citation networks uses a random walker on a causal set to estimate it's spectral dimension\cite{Eichhorn2014}.
However many other methods are inappropriate for analysis of the causal structure of a citation network. One such method is to find the smallest dimension in which any subgraph can be faithfully embedded \cite{Bombelli1987a, Brightwell1991a}. This method requires a DAG to be perfectly embedded in a manifold as just one subgraph which cannot be embeddable in $D$ dimensions means that the entire network is given a dimension higher than $D$. 

Furthermore, there are some finite DAGs which cannot be perfectly embedded in a Minkowski space of any dimension \cite{Felsner1999}.
These methods are less appropriate for analysis of our citation networks, since the (integer) result such dimension estimates give can be increased by one by the rewiring of only one edge, which is an unhelpful property when dealing with noisy real-world data, or it may not even be defined at all.

Conversely the two estimates we use here are robust to noise, a useful property when analysing data from social interactions. They produce a real number value and and small deviations from DAGs which are faithfully embedded in a Minkowski manifold only lead to small deviations in the estimated dimension.

\section*{Acknowledgments}

We would like to thank Jamie Gollings, Amy Hughes, Tamar Loach, Kim Christensen and Mauricio Barahona for useful discussions.

\appendix
\section{Reduced Degree}
\subsection*{Derivation of distribution of reduced degree in 2D spacetime networks}
In 2D Minkowski space the exact distribution of degrees after TR can be calculated because in 2 dimensions the spacetime network is equivalent to another structure called a cube space. In a cube space of dimension $D$ we have points $i=1,2...N$ with coordinates $z_i^\alpha$ where $\alpha=1,2...D$.
Point $i$ is connected to point $j$ iff $z_i^\alpha < z_j^\alpha \; \forall \; \alpha$, which is to say that $j$ has larger co-ordinates in all dimensions. The nodes of the network are randomly and uniformly scattered in this space and connected using this rule.

Though this result will turn out to be of little help for analysis of citation networks, for reasons discussed later, it is to the best of our knowledge a novel extension of the result of \cite{Winkler1985} which derives the expected value of the reduced degree in cube spaces of all dimensions and so we include it here.

Suppose that the probability of a node in the corner of an interval containing $N$ other nodes, having a degree after TR (reduced degree) $\kred$ is $p(\kred, N)$.
We will first give an argument for the following recursion relation.
\begin{equation}
p(\kred, N) = \frac{N-1}{N} p(\kred, N-1) + \frac{1}{N} p(\kred - 1, N-1)
\label{eq:recursion}
\end{equation}
Since we are only considering 2 dimensions, let us call the first coordinate $x$, $z_i^{\alpha=1} = x_i$, and the second coordinate $y$, $z_i^{\alpha=2} = y_i$. We may consider each point in turn, ordering them with largest $x$ coordinates first so that $x_i > x_j$ if $i<j$. Suppose we have already considered the first $(N-1)$ points and now look at the point with the $N$-th largest $x$ coordinate, $i=N$. This point $i=N$ can only be a new link to the origin if it is minimal in the $y$ coordinate.  That is, since we already know that every existing point has a larger $x$ coordinate by our ordering, we have $x_i>x_j$ but $y_i<y_j$ for $N=i>j$. Because the coordinates are just random numbers, the probability that $y_N$ is the smallest is simply $\frac{1}{N}$. So with this probability, a new TR-surviving-edge will appear, and with probability $\frac{N}{N-1}$ it will not, explaining both terms in equation ~\ref{eq:recursion}. This view is equivalent to a standard record statistics process \cite{SJ13}.
Indeed the points don't even have to be uniformly distributed here, the only requirement is that the that the $D$ coordinates are independent random variables. 

To solve this, we then recognise the recursion relation for the \href{http://mathworld.wolfram.com/StirlingNumberoftheFirstKind.html}{unsigned Stirling numbers of the first kind} 
$\left[ \begin{array}{c} N \\ {\kred} \end{array}  \right]$, namely

\begin{eqnarray}
\left[ \begin{array}{c} N+1 \\ {\kred} \end{array}  \right] &=&
N \left[ \begin{array}{c} N \\ {\kred} \end{array}  \right] +
\left[ \begin{array}{c} N \\ {\kred - 1} \end{array}  \right]
\\
\mbox{where} \left[ \begin{array}{c} 0 \\ 0 \end{array}  \right] &=& 1
\\
 \mbox{and}
\left[ \begin{array}{c} N \\ 0 \end{array}  \right] &=& \left[ \begin{array}{c} 0 \\ N \end{array}  \right] = 0
\label{eq:stirling1}
\end{eqnarray}

We can then say that
\begin{equation}
p(\kred, N) = \frac{1}{N!} \left[ \begin{array}{c} N \\ {\kred} \end{array}  \right]
\label{eq:stirling2}
\end{equation}
To check our answer, note that $\left[ \begin{array}{c} N \\ 1 \end{array} \right] = (N-1)!$ giving $p(\kred=1, N) = \frac{1}{N}$ as expected
\footnote{The probability that $\kred=1$ is simply the probability that the point with the smallest x-coordinate also has the smallest y-coordinate}.
As noted by Wilf in \cite{Wilf1993}, `the Stirling numbers of the first kind are notoriously difficult to compute', and so we are unlikely to find a nice solution here. 

It is useful to find the generating function $G(z, N)$ where
\begin{equation}
G(z, N) = \sum^{\infty}_{k=0} z^k p(k, N)
\label{eq:generating_function}
\end{equation}
with $p(\kred, N) = 0$ if $\kred>N$. Note that $G(z=1, N) = 1$ and the first term in this polynomial is $\frac{z}{N}$ because $p(k=0, N)=0$. From the recursion relation ~\ref{eq:recursion} we now find that
\begin{eqnarray}
G(z, N) & = &  \frac{N-1}{N} G(z, N-1) + \frac{z}{N}G(z, N-1)
\label{eq:p_gen_func_1}
\\
G(z, N) &=& \frac{\Gamma(N+z)}{\Gamma(z) \Gamma(N+1)} = (z+N-1).(z+N-2)...(z) \; \times \frac{1}{N!}  \, .
\label{eq:ratio_of_gammas}
\end{eqnarray}
Note that the $\Gamma(z)$ normalisation factor on the denominator can be seen from the explicit expansion where we know the term $O(z)$ is $\frac{z}{N}$. 

The asymptotic limit \cite{Wilf1993,Hwang1995} can be studied from the generating function $G(z, N)$ in ~\ref{eq:ratio_of_gammas} as
\begin{eqnarray}
\lim\limits_{N \to \infty} G(z, N) &=& \frac{N^{z-1}}{\Gamma(z)}
\label{eq:limits_1}
\\
&=& \frac{1}{N} \frac{\sum_{k=0} (\ln(N))^k z^k / k!}{z^{-1} + \psi(0) + O(z)}
\label{eq:limits_2}
\end{eqnarray}
The first term in the series, the part coming from the $\Gamma(N+z)/\Gamma(z)$ is just the generating function for the Poisson distribution $p_{\mathrm{Poisson}}(k) = e^{-\lambda} \lambda^k / k!$ with mean $\lambda = \ln(N)$, divided by $\Gamma(z)$. However, the non-leading terms coming from the  expansion of the denominator, $ \Gamma(z)$, prevent a simple match so the Poisson-like behaviour as seen in \cite{Bolognesi2014} may only be useful for small ranges of $\kred$, typically $|\Delta \kred| \ll \ln(N)$.

From the generating function in ~\ref{eq:ratio_of_gammas} we can find various moments of $\kred$ for fixed $N$. Here we will derive the expected $\kred$ for a given $N$ in two-dimensions.
\begin{eqnarray}
\langle \kred \rangle = \sum_{\kred=0}^\infty \kred \, p(\kred, N)
&=& \left. \frac{\partial G(z, N)}{\partial z} \right| _{z=1}
= \sum_{i=1}^N \frac{1}{i} = H_n \approx \gamma + \ln(N) \, ,
\end{eqnarray}
where  $\gamma \approx 0.577$ is the Euler-Mascheroni constant.
This Harmonic number result is the two-dimensional case of $D$-dimensional result in \cite{Winkler1985}, which in the large $N$ limit tends to logarithmic growth, as suggested in \cite{Eichhorn2014}.

%\newpage
%
%\iffalse
%Using the asymptotic expansions in theorem 2, equation 5 of \cite{Hwang1995} we find
%\begin{equation}
%p(\kred, N) \approx \frac{1}{N} \frac{(\ln N)^{\kred -1}}{(\kred -1)!} \left( \frac{1}{\Gamma(1+r)} + something\right)
%\label{eq:asymptotic}
%\end{equation}
%
%\begin{equation}
%r = \frac{(\kred -1)}{\ln(N)},\quad 2 \leq \kred \leq \eta \ln(N),\quad \eta > 0
%\label{eq:r}
%\end{equation}
%\fi

\subsection*{Comparison of measured reduced degree in citation networks, and spacetime networks}
\begin{figure}[!h]
\centering
\includegraphics[width=0.49\textwidth]{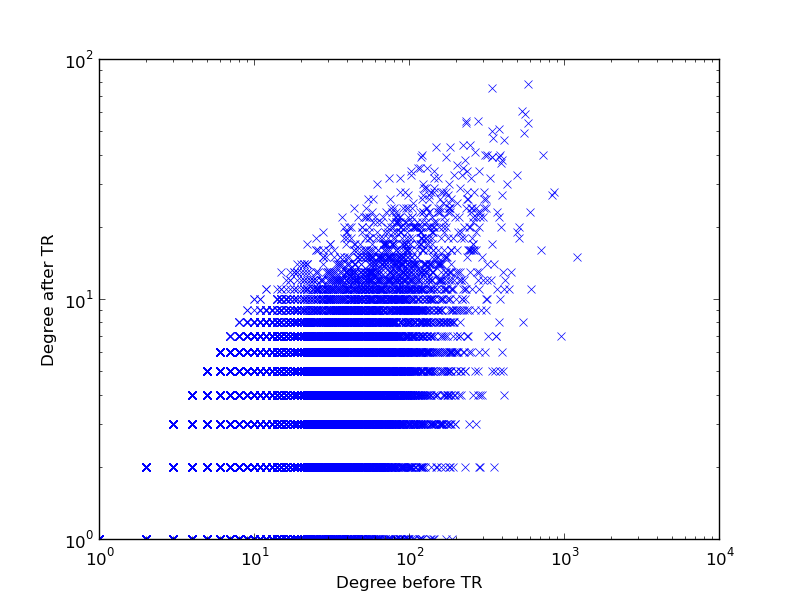}%}
\includegraphics[width=0.49\textwidth]{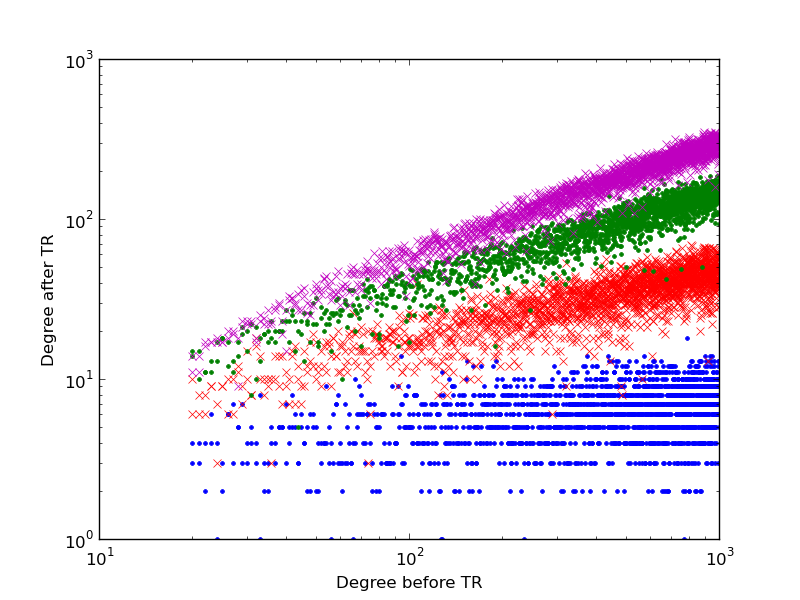}%}
\caption{\textbf{Left:} the degree before, and after transitive reduction for the \texttt{hep-ph} citation network. The spread of $\kred$ is very wide for a given $\kred$, indicating a heterogeneity in the papers.\newline
\textbf{Right:} the degree before, and after transitive reduction for spacetime networks of dimension 2-5. Lower dimension appear lower on the plot.\newline
To try and use the reduced degree method to estimate dimension is essentially to ask which of the scatter plots on the right figure best fits the left figure, given the large spread of values on the left, the estimated dimension for individual subgraphs has very large variation, unlike the other dimension estimates which have similar answers throughout the network and so better achieve the goal of characterising the whole network's structure. The reduced degree method is more useful as a characterisation of individual nodes within the network.}
\label{fig:reduced_degree}
\end{figure}

\end{document}